\pdfoutput=1
\documentclass[prl,twocolumn,showpacs,amsmath,amssymb,floatfix,superscriptaddress]{revtex4-1}

\usepackage{graphicx}
\usepackage{dcolumn}
\usepackage{bm}
\usepackage[colorlinks=true, linkcolor=blue, urlcolor=blue,citecolor=blue]{hyperref}

\DeclareMathOperator{\sech}{sech}

\begin{document}
	
\title{Wide ferromagnetic domain walls can host both adiabatic reflectionless spin transport and finite nonadiabatic spin torque:  A time-dependent quantum transport picture}

		
		
\author{Felipe Reyes Osorio}
\affiliation{Department of Physics and Astronomy, University of Delaware, Newark, DE 19716, USA}
\affiliation{Departamento de F\'{i}sica, Universidad del Valle, AA 25360, Cali, Colombia}
\author{Branislav K. Nikoli\'{c}}
\email{bnikolic@udel.edu}
\affiliation{Department of Physics and Astronomy, University of Delaware, Newark, DE 19716, USA}
		
		
\begin{abstract}
	The key concept in spintronics of current-driven noncollinear magnetic textures, such as magnetic domain walls (DWs), is  {\em adiabaticity}, i.e., how closely electronic spins track classical localized magnetic moments (LMMs) of the texture.  When mistracking occurs nonadiabatic effects arise, the salient of which is  {\em nonadiabatic} spin transfer torque (STT) where spin angular momentum is exchanged  between  electrons and LMMs to cause their dynamics and enable DW motion without any current threshold. The microscopic mechanisms behind nonadiabatic STT have been debated theoretically for nearly two decades, but with unanimous conclusion that they should be significant only in {\em narrow} DWs. However, this contradicts {\em sharply}  experiments [O. Boulle {\em et al.}, Phys. Rev. Lett. {\bf 101}, 216601 (2008); C. Burrowes {\em et al.}, Nat. Phys. {\bf 6}, 17 (2010)] observing nonadiabatic STT in DWs much wider than putatively relevant \mbox{$\sim 1$ nm} scale, as well as largely insensitive to further increasing the DW width $w$. Here we employ time-dependent quantum transport for electrons, combined self-consistently with  the Landau-Lifshitz-Gilbert (LLG) equation for LMMs,  to obtain both nonadiabatic and adiabatic STT from  the {\em exact} nonequilibrium density matrix and its lowest order  as adiabatic density matrix defined by assuming that LMMs are infinitely slow. This allows us to demonstrate that our microscopically, and without any simplifications of prior derivations like effectively static DW, extracted  nonadiabatic STT: ({\em i}) does not decay, but instead {\em saturates} at a finite value, with increasing $w$ of a moving DW ensuring entry into the adiabatic limit, which we characterize by showing that electronic spins do not reflect from the static DW in this limit; and ({\em ii}) it has both out-of-DW-plane, as is the case of phenomenological expression widely used in the LLG equation, and in-plane components, where the former remains finite with increasing $w$.
\end{abstract}

\maketitle

\begin{figure}
	\centering
	\includegraphics[scale=0.14]{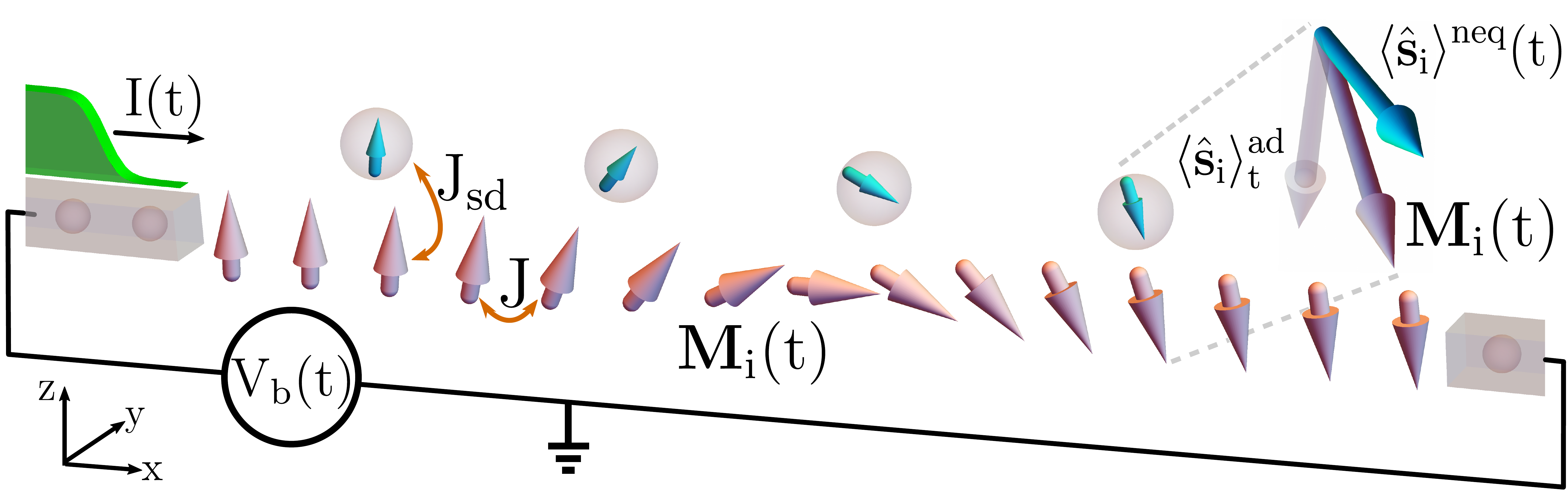}
	\caption{Schematic view of FM nanowire of $N=40$ sites hosting a DW and sandwiched between two semi-infinite NM leads. Bias voltage $V_b(t)$ is turned on at $t=0$ to inject unpolarized charge current from the left NM lead. Light red arrows illustrate orientation of unit vectors $\mathbf{M}_i$ of classical LMMs comprising two domains and DW in between; gray and cyan arrows illustrate adiabatic  $\langle \hat{\mathbf{s}}_i \rangle^\mathrm{ad}_t$ [Eq.~\eqref{eq:sadiabatic}] and nonequilibrium $\langle \hat{\mathbf{s}}_i \rangle^\mathrm{neq}(t)$ [Eq.~\eqref{eq:snonadiabatic}] expectation values of electron spin, respectively. These three vectors are in general {\em noncollinear}, so that cross products of spin expectation values with $\mathbf{M}_i$ define nonadiabatic [Eq.~\eqref{eq:nadstt}] and adiabatic [Eq.~\eqref{eq:adstt}] STTs.}
	\label{fig:fig1}
\end{figure}

{\em Introduction}.---One of the key concept in electronic spin transport through noncollinear static or dynamic (i.e., time-dependent) magnetic textures---such as magnetic domain walls (DWs)~\cite{Tatara2008,Tatara2019,Boulle2008,Burrowes2010}, skyrmions~\cite{Akosa2017} and vortex cores~\cite{Heyne2010,Pollard2012}---is that of {\em adiabaticity}~\cite{Tatara2019,Levy1997,Zhang2004,Barnes2005,Gregg1996,Berger1978,Aharonov1992,Xiao2006}. For example, within a ferromagnetic nanostructure, DW is a transition region that separates two different but uniformly magnetized regions, as  illustrated in Fig.~\ref{fig:fig1}. Moving DWs with charge currents~\cite{Kim2017a}, instead of external magnetic fields, is a topic of both great fundamental interest~\cite{Tatara2008,Tatara2019} for nonequilibrium quantum many-body physics and applications of DWs for digital memory~\cite{Parkin2015} and logic~\cite{Allwood2002} or neuromorphic computing~\cite{Grollier2016}. 

The DWs in nanowires of conventional $3d$ metallic ferromagnets (Fe, Ni, Co) can be thick (i.e., of width \mbox{$w \sim 100$ nm}) and $sd$ exchange interaction $J_\mathrm{sd}$  between the spin of conduction electrons and localized spins is strong~\cite{Cooper1967}, so that intuitive picture~\cite{Tatara2019,Levy1997,Gregg1996} is the one in which electronic spin tracks spatially noncollinear magnetic textures while traversing DW. In clean magnetic wires, the condition~\cite{Tatara2019,Waintal2004,Barnes2005} for such {\em adiabatic} limit is 
\begin{equation}\label{eq:clean}
J_\mathrm{sd} w S/\hbar v_F \gg 1.
\end{equation}
This ``adiabaticity parameter''  is defined as the ratio of two time scales---$w/v_F$ needed for electron to traverse the DW of width $w$ with Fermi velocity $v_F$, and $\hbar/J_\mathrm{sd} S$ governing electron spin rotation within the DW (composed of localized spins $S$) by $sd$ exchange interaction. This criterion can be generalized~\cite{Ban2009} to $J_\mathrm{sd} w^2/\hbar \mathcal{D} \gg 1$ for diffusive electronic transport ($\mathcal{D}$ is the diffusion constant). Since $J_\mathrm{sd}$ (typically measured as \mbox{$\sim 0.1$ eV}~\cite{Cooper1967}), $v_F$ and $\mathcal{D}$ are fixed by materials properties, increasing $w$ can satisfy either of these two conditions to explain how wide DW allows electron to pass through it without reflection~\cite{Tatara2019} and, therefore, with vanishing electrical resistance of DW. 

In spintronic experiments and applications, any deviation from adiabaticity of electron spin dynamics leads to fundamental effects, such as finite DW resistance~\cite{Levy1997,Tatara1997,Tatara2000,Simanek2001,Simanek2005,Berger2007,Gopar2004,Tatara2019} and particularly important {\em nonadiabatic}~\cite{Zhang2004,Thiaville2005} spin-transfer torque (STT)  in electronic transport through DWs~\cite{Boulle2008,Burrowes2010}, skyrmions~\cite{Akosa2017} and vortex cores~\cite{Heyne2010,Pollard2012}. The STT is a phenomenon~\cite{Ralph2008} in which flowing electrons transfer spin angular momentum to local magnetization $\mathbf{M}(\mathbf{r})$ viewed as classical vector, as long as nonequilibrium spin expectation value of an electron and $\mathbf{M}(\mathbf{r})$ are noncollinear, as illustrated in Fig.~\ref{fig:fig1}. Spintronic experiments on current-driven dynamics of noncollinear textures are standardly interpreted~\cite{Boulle2008,Heyne2010,Burrowes2010,Pollard2012} using the Landau-Lifshitz-Gilbert (LLG) equation 
\begin{eqnarray}\label{eq:llg}
	\partial_t \mathbf{M} & = &  - g_0 \mathbf{M} \times \mathbf{B}_\mathrm{eff} + \underbrace{\lambda \mathbf{M} \times \partial_t \mathbf{M}}_\text{Gilbert damping} \nonumber \\ 
	&& - \underbrace{(\mathbf{u} \cdot \nabla) \mathbf{M}}_\text{adiabatic STT} + \underbrace{\beta \mathbf{u} \mathbf{M} \times \big[(\mathbf{u} \cdot \nabla) \mathbf{M} \big]}_\text{nonadiabatic STT},
\end{eqnarray}
extended to include adiabatic~\cite{Bazaliy1998,Tatara2004,Fernandez-Rossier2004} and nonadiabatic~\cite{Zhang2004,Thiaville2005,Cheng2013} STT terms. Here we use  shorthand notation $\partial_t \equiv \partial/\partial t$; $g_0$ is gyromagnetic factor~\cite{Evans2014}; $\mathbf{B}_\mathrm{eff}$ is the effective magnetic field;  $\lambda$ is dimensionless Gilbert damping~\cite{Evans2014}; and velocity $\mathbf{u}$ is proportional~\cite{Burrowes2010} to current density $\mathbf{j}$ and its polarization $P$ acquired in passing through the domains. While adiabatic STT is well understood and has been reproduced by a number of different transport theories~\cite{Duine2007,Kishine2010,Mondal2018,Xiao2006}, the nonadiabatic STT in Eq.~\eqref{eq:llg} has provoked a lively debate over the past (nearly) two decades about its possible mechanisms, their magnitude and the ratio $\beta/\lambda$~\cite{Duine2007,Barnes2005}. For example, the contributions discussed include spin relaxation, due to spin-flip scattering off spin-orbit (SO) or magnetic impurities, acting as an effective nonadiabaticity~\cite{Zhang2004,Kohno2006,Tatara2008a,Cheng2013}; momentum transfer~\cite{Tatara2004,Tatara2008}; genuine spin-mistracking~\cite{Waintal2004} and relativistic effects~\cite{Mondal2018}. Also, even in adiabatic limit specified by Eq.~\eqref{eq:clean} strong intrinsic (i.e., from band structure) SO coupling can generate nonzero $\beta$-term~\cite{Nguyen2007,Garate2009}. The understanding of conditions for nonzero $\beta$-term  is {\em crucial} for anticipated DW motion-based technologies  because nonadiabatic STT makes possible current-driven DW motion at any finite current and in the absence of externally applied magnetic field.

In particular, microscopic quantum-mechanical calculations~\cite{Cheng2013,Duine2007,Kohno2006,Tatara2008a}, going beyond early phenomenological analysis~\cite{Zhang2004,Thiaville2005}, have focused on computing the dimensionless $\beta$ parameter while not questioning  the form of 
nonadiabatic STT in Eq.~\eqref{eq:llg}. An {\em exception} is Ref.~\cite{Xiao2006} where different form has been proposed, with both in-plane and out-of-plane components of nonadiabatic STT (see Eq.~(34) in Ref.~\cite{Xiao2006}), unlike the one in Eq.~\eqref{eq:llg} which has only~\cite{Tatara2019,Zhang2004,Kishine2010} out-of-DW-plane (assuming N\'{e}eel DW) component. Nevertheless, both standard nonadiabatic STT and modified form of Ref.~\cite{Xiao2006} decay (such as, $\propto w^{-1}e^{-cw/\ell}$ in Ref.~\cite{Xiao2006} or $\propto 1/w$ in Ref.~\cite{Waintal2004}) as the DW width $w$ increases due to diminishing magnetization gradient in $(\mathbf{u} \cdot \nabla) \mathbf{M}$. Thus, unless $w$ is comparable to relevant transport scale $\ell$---as set by the Fermi wavelength, Larmor precession or the mean free path, being of the order of \mbox{$\sim 1$ nm} in transition ferromagnetic metals (FMs)---nonadiabatic STT is negligible. In {\em sharp contrast}, variety of experimental techniques developed to directly measure $\beta$ have observed large nonadiabatic STT in DW of much larger width \mbox{$w \sim 10$ nm}~\cite{Boulle2008,Burrowes2010,Heyne2010}. This suggests that widely used nonadiabatic STT term in Eq.~\eqref{eq:llg} {\em does not} fully capture all relevant nonadiabatic effects in electronic spin transport through noncollinear magnetic textures. 

\begin{figure}
	\includegraphics[scale=0.26,angle=0]{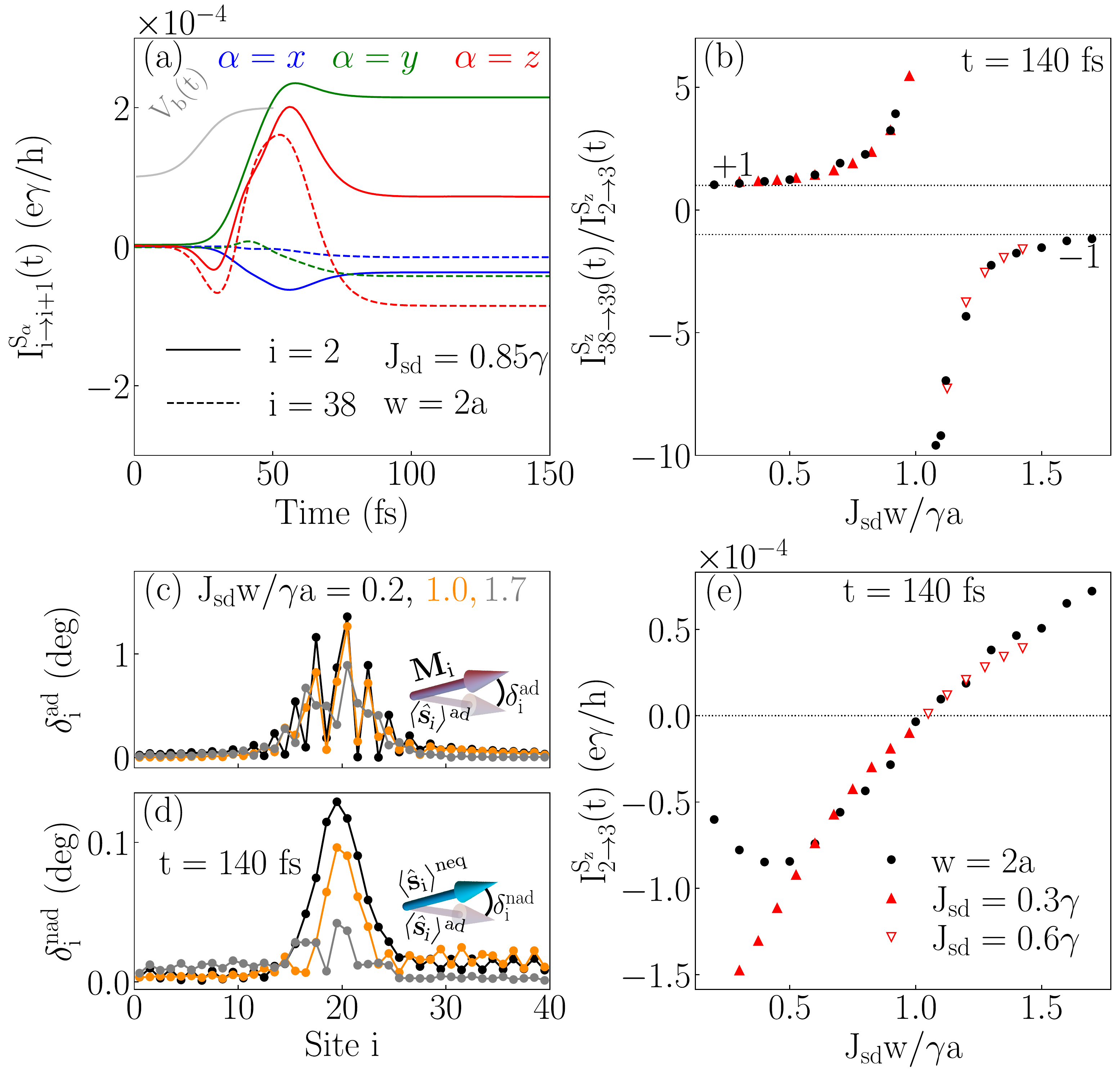}
	\caption{(a) Time dependence of local (or bond~\cite{Nikolic2006,Petrovic2018}) spin currents $I^{S_\alpha}_{i \rightarrow i+1}(t)$ from site $i$ to site $i+1$ before ($i=2$) and after  ($i=38$) DW kept static in Fig.~\ref{fig:fig1}. These currents are driven by injecting unpolarized charge current by bias voltage $V_b(t)$ with profile shown as gray curve. (b) Ratio of these two local currents for $S_\alpha \equiv S_z$  at $t=140$ fs, when steady state transport regime is reached in panel (a), for different values of adiabaticity parameter in Eq.~\eqref{eq:clean} increased by changing $J_\mathrm{sd}$ at fixed DW width $w$ (black circles) or vice versa (red triangles). The divergence in panel (b) around $J_\mathrm{sd} w/\gamma a \simeq 1.0$ is due to $I^{S_z}_{2 \rightarrow 3} \rightarrow 0$ in panel (e).  Spatial profile of angles (see insets for definitions) between vectors [Fig.~\ref{fig:fig1}] of: (c) adiabatic electronic spin density [Eq.~\eqref{eq:sadiabatic}] and $\mathbf{M}_i$; and (d) adiabatic and nonequilibrium [Eq.~\eqref{eq:totalstt}] electronic spin densities.}
	\label{fig:fig2}
\end{figure}

\begin{figure}
	\includegraphics[scale=0.26,angle=0]{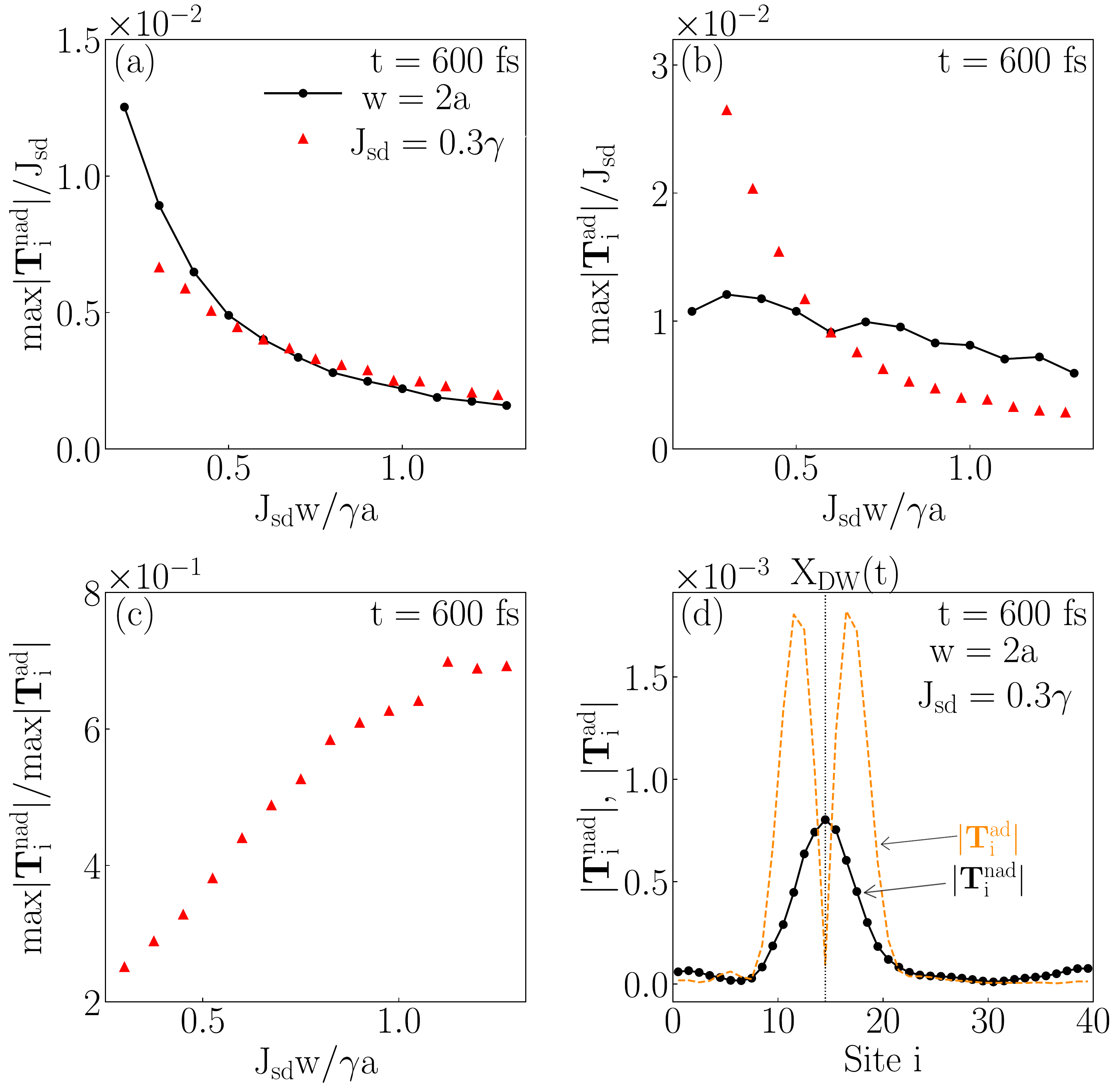}
	\caption{(d) Spatial dependence of the magnitude of nonadiabatic $|\mathbf{T}^\mathrm{nad}_i(t)|$  [Eq.~\eqref{eq:nadstt}] and  adiabatic  $|\mathbf{T}^\mathrm{ad}_i(t)|$  [Eq.~\eqref{eq:adstt}] STT at $t=600$ fs, after bias voltage \mbox{$V_b=0.05$ eV} is turned on (as a step function) at $t=0$ to inject unpolarized charge current in  Fig.~\ref{fig:fig1}. Maximum values from panel (d) are plotted in panels (a) and (b) for nonadiabatic and adiabatic STT, respectively, with their ratio shown in panel (c),  as a function of adiabaticity parameter [Eq.~\eqref{eq:clean} and Fig.~\ref{fig:fig2}] varied by increasing the DW width $w$ (triangles) or $J_\mathrm{sd}$ (circles + solid line).}
	\label{fig:fig3} 
\end{figure}

\begin{figure}
	\includegraphics[scale=0.26]{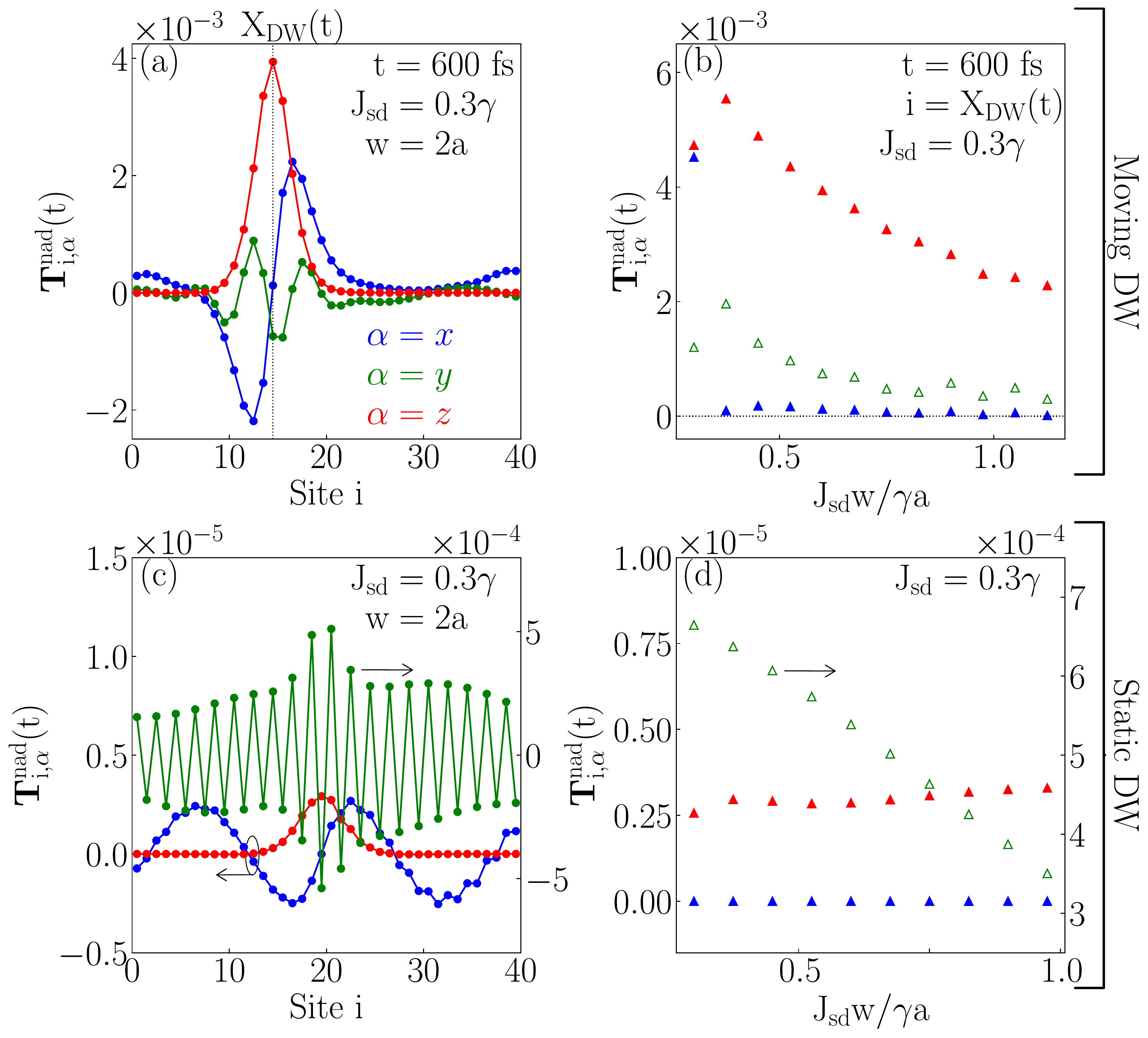}
	\caption{Spatial profile of the Cartesian components of nonadiabatic STT [Fig.~\ref{fig:fig3}(d)] for: (a) moving DW due to injected unpolarized electronic current; or (c) static DW [i.e., LLG Eq.~\eqref{eq:llglattice} is coupled to TDNEGF in (a) but not in (c)]. Panels (b) and (d) show nonadiabatic STT components from (a) and (b), respectively, at the center $X_\mathrm{DW}$ of DW,  as a function of adiabaticity parameter   [Eq.~\eqref{eq:clean} and Fig.~\ref{fig:fig2}] varied by increasing the DW width $w$.}
	\label{fig:fig4} 
\end{figure}

In this Letter, we employ numerically {\em exact} and fully microscopic (i.e., requiring only quantum Hamiltonian of electrons and classical Hamiltonian of LMMs) time-dependent quantum transport (QT) formalism~\cite{Gaury2014,Popescu2016,Petrovic2018,Bajpai2019a,Bajpai2020,Petrovic2021}  to resolve this issue. This is achieved by splitting the exact nonequilibrium density matrix~\cite{Bajpai2020,Gaury2014,Popescu2016} of time-dependent QT into two terms to rigorously define adiabatic and nonadiabatic STT, whose properties are then studied as a function of DW width $w$. Our principal results in Fig.~\ref{fig:fig3} and ~\ref{fig:fig4} show that thus defined nonadiabatic STT becomes insensitive to increasing $w$, once putative adiabatic limit is reached as signified~\cite{Tatara2019} by electronic spins {\em ceasing} to reflect from DW [Fig.~\ref{fig:fig2}(b)]. Thus, our  predictions are in full accord with apparently highly surprising experimental observations~\cite{Boulle2008,Heyne2010,Burrowes2010,Pollard2012}. Prior to delving into these results, we first introduce rigorous definitions of adiabatic and nonadiabatic electronic spin expectation values---also visualized in Figs.~\ref{fig:fig1}, ~\ref{fig:fig2}(c) and ~\ref{fig:fig2}(d)---and the corresponding STTs, as well as our models and time-dependent QT methodology.

{\em Defining adiabaticity and and nonadiabatic STT rigorously from time-dependent QT}.---The intuitive picture (see, e.g., Figs. 14 and 19 in Ref.~\cite{Tatara2019} or Fig. 4 in Ref.~\cite{Gregg1996}) of how spin of injected conduction electron propagates through a magnetic DW is commonly  built around the idea that in the adiabatic limit expectation value of electron spin will be parallel to local magnetization. This is also a usual starting point of phenomenological calculations of nonadiabatic STT  where one assumes that nonequilibrium electronic spin density can be split as $\langle \hat{\mathbf{s}} (\mathbf{r}) \rangle^\mathrm{neq}(t) = \langle \hat{\mathbf{s}}(\mathbf{r}) \rangle^\mathrm{ad}(t) + \delta \langle \hat{\mathbf{s}} (\mathbf{r}) \rangle(t)$ (see, e.g., Eq. (5) in Ref.~\cite{Zhang2004}). Here $\langle \hat{\mathbf{s}} (\mathbf{r}) \rangle^\mathrm{ad}(t) \parallel \mathbf{M}(\mathbf{r})$ is assumed~\cite{Zhang2004} to point along  $\mathbf{M}(\mathbf{r})$ and represents adiabatic spin density  due to  electronic spin relaxing to its {\em equilibrium} value at an instantaneous  time $t$;  and $\delta \langle \hat{\mathbf{s}}(\mathbf{r}) \rangle (t)$ is nonadiabatic correction. 

However, in spintronics~\cite{Berger1978}, Berry phase~\cite{Aharonov1992,Stahl2017} and noncollinear density functional theory~\cite{Capelle2001} literature it has been already noticed that electronic spin {\em never} [Figs.~\ref{fig:fig2}(c) and ~\ref{fig:fig2}(d)] simply aligns with $\mathbf{M}(\mathbf{r})$. Instead, $\langle \hat{\mathbf{s}} (\mathbf{r}) \rangle^\mathrm{ad}(t)$ tends to align perfectly with another effective~\cite{Aharonov1992} local magnetization \mbox{$\mathbf{M}_\mathrm{eff}(x) = \mathbf{M}(x) + Q(v_F) d\mathbf{M}(x)/dx \times \mathbf{M}(x)$}, where $Q$ is a function of electron velocity~\cite{Xiao2006} and we assume sufficiently slowly varying $\mathbf{M}(x)$ along the $x$-axis (as pertinent to Fig.~\ref{fig:fig1}). This also means that the natural basis for an electron spin moving through a noncollinear magnetic texture 
is not along the local magnetization $\mathbf{M}(x)$, as often assumed~\cite{Levy1997,Zhang2004,Tatara2019}, but rather along $\mathbf{M}_\mathrm{eff}(x)$. Thus, even in equilibrium~\cite{Capelle2001}, or in nonequilibrium but in perfectly adiabatic~\cite{Stahl2017,Bajpai2020} limit, there is nonzero~\cite{Stahl2017,Bajpai2020,Capelle2001} local torque $\langle \hat{\mathbf{s}} (\mathbf{r}) \rangle^\mathrm{ad}(t) \times \mathbf{M}(\mathbf{r},t) \neq 0$.

Since ``adiabatic'' means that electron spin remains in the the lowest energy state at each time, in closed quantum systems~\cite{Stahl2017} adiabatic spin density at time $t$ can be computed as the expectation value \mbox{$\langle \hat{\mathbf{s}}_i \rangle^\mathrm{ad}_t = \langle \Psi_0(t)| \big(|i\rangle \langle i| \otimes\hat{\bm \sigma}\big)| \Psi_0(t) \rangle$} in the  instantaneous ground state $|\Psi_0\rangle$ of the Hamiltonian $\hat{H}[\mathbf{M}_i(t)]$ for configuration of LMMs at time $t$, \mbox{$|\Psi(t)\rangle = |\Psi_0[\mathbf{M}_i(t)]\rangle$}. Here  we switch the notation from  $\mathbf{M}(\mathbf{r})$  to  unit vectors $\mathbf{M}_i$ describing LMMs on the sites of a discrete lattice in Fig.~\ref{fig:fig1}, hosting both them and localized electronic orbitals $|i\rangle$. In an open quantum system, such as the one in Fig.~\ref{fig:fig1}, we need  adiabatic~\cite{Bajpai2020,Bode2011,Thomas2012a} density matrix ${\bm \rho}^\mathrm{ad}_t$  to express $\langle \hat{\mathbf{s}} \rangle^\mathrm{ad}_t$
\begin{eqnarray}
	{\bm \rho}^\mathrm{ad}_t & = & -\frac{1}{\pi} \int dE\, \mathrm{Im} \mathbf{G}^r_t f(E), \label{eq:rhoadiabatic} \\
	\langle \hat{\mathbf{s}}_i \rangle^\mathrm{ad}_t & = & \mathrm{Tr}\, \big[{\bm \rho}^\mathrm{ad}_t | i\rangle \langle i| \otimes \hat{\bm \sigma} \big]. \label{eq:sadiabatic}
\end{eqnarray}
Here \mbox{$\mathbf{G}^r_t=\big[E - \hat{H}[\mathbf{M}_i(t)] - {\bm \Sigma}_L - {\bm \Sigma}_R \big]^{-1}$} is the retarded Green's function (GF); ${\bm \Sigma}_{L,R}$ are self-energies due to the left ($L$) and right ($R$) NM leads; and \mbox{$\mathrm{Im} \mathbf{G}^r_t=(\mathbf{G}^r_t - [\mathbf{G}^r_t]^\dagger)/2i$}. Such ${\bm \rho}^\mathrm{ad}_t$  is grand canonical equilibrium density matrix, but depending parametrically~\cite{Bode2011,Thomas2012a,Mahfouzi2016} (or implicitly, so we put $t$ in the subscript) on time via instantaneous configuration of $\mathbf{M}_i(t)$, thereby effectively assuming \mbox{$\partial_t  \mathbf{M}_i(t)= 0$}. 

The nonadiabatic corrections take into account \mbox{$\partial_t  \mathbf{M}_i(t) \neq 0$}, so that ${\bm \rho}^\mathrm{ad}_t$ can be viewed as the lowest order term~\cite{Bajpai2020,Bode2011,Thomas2012a} of nonequilibrium density matrix ${\bm \rho}^{\rm neq}(t)$ expanded~\cite{Mahfouzi2016} in powers of small  $\partial_t  \mathbf{M}_i(t)$. We compute  ${\bm \rho}^{\rm neq}(t) = \hbar \mathbf{G}^<(t,t)/i$ exactly in terms of  the lesser Green's function of time-dependent nonequilibrium GF (TDNEGF) formalism~\cite{Stefanucci2013,Gaury2014}, which makes it possible to obtain nonadiabatic spin density $\delta \langle \hat{\mathbf{s}}_i  \rangle (t)$ and the corresponding nonadiabatic STT as 
\begin{eqnarray}
	\delta \langle \hat{\mathbf{s}}_i  \rangle (t) & = &\langle \hat{\mathbf{s}}_i  \rangle^\mathrm{neq} (t) - \langle \hat{\mathbf{s}}_i  \rangle^\mathrm{ad}_t \nonumber \\ \mbox{} &=&  \mathrm{Tr}\, \big[({\bm \rho^{\rm neq}}(t) - {\bm \rho}^\mathrm{ad}_t) | i\rangle \langle i| \otimes \hat{\bm \sigma}\big], \label{eq:snonadiabatic} \\
	\mathbf{T}^\mathrm{nad}_i(t) & = & J_\mathrm{sd} \, \delta \langle \hat{\mathbf{s}}_i \rangle(t) \times \mathbf{M}_i(t), \label{eq:nadstt}
\end{eqnarray}
while adiabatic STT is given by
\begin{equation}\label{eq:adstt}
	\mathbf{T}^\mathrm{ad}_i(t) =   J_\mathrm{sd} \langle \hat{\mathbf{s}}_i \rangle^\mathrm{ad}_t \times \mathbf{M}_i(t).
\end{equation}
The total STT acting on DW is then given by 
\begin{eqnarray}\label{eq:totalstt}
\mathbf{T}_i & = &  \mathbf{T}^\mathrm{ad}_i + \mathbf{T}^\mathrm{nad}_i \nonumber \\
& = &  J_\mathrm{sd} \langle \hat{\mathbf{s}}_i \rangle^\mathrm{neq} \times \mathbf{M}_i \nonumber \\
& = &  J_\mathrm{sd}\mathrm{Tr}\, \big[{\bm \rho^{\rm neq}} | i\rangle \langle i| \otimes \hat{\bm \sigma}\big] \times \mathbf{M}_i,
\end{eqnarray}
where the distinction between  $\langle \hat{\mathbf{s}}_i \rangle^\mathrm{neq}(t)$ and $\langle \hat{\mathbf{s}}_i \rangle^\mathrm{ad}_t$, as well as their orientations with respect to $\mathbf{M}_i$, are illustrated in Fig.~\ref{fig:fig1} and quantified in Fig.~\ref{fig:fig2}. For example,  Fig.~\ref{fig:fig2}(d) shows that increasing $J_\mathrm{sd}$ or $w$ tends to align  $\langle \hat{\mathbf{s}}_i \rangle^\mathrm{neq}(t)$ with $\langle \hat{\mathbf{s}}_i \rangle^\mathrm{ad}_t$, which also justifies that Eqs.~\eqref{eq:rhoadiabatic} and ~\eqref{eq:sadiabatic} are proper description of spin adiabaticity. But  $\langle \hat{\mathbf{s}}_i \rangle^\mathrm{ad}_t$ remains misaligned with $\mathbf{M}_i$ in Fig.~\ref{fig:fig2}(c), in accord with prior analysis of Refs.~\cite{Berger1978,Aharonov1992,Xiao2006}.

We underscore that $\delta \langle \hat{\mathbf{s}}(\mathbf{r}) \rangle$ was originally obtained~\cite{Zhang2004} from semiclassical transport theory, leading to phenomenological expression for nonadiabatic STT in Eq.~\eqref{eq:llg}, while assuming~\cite{Zhang2004,Tatara2019} that ``$\delta$'' denotes deviation from $\mathbf{M}(\mathbf{r})$ rather than from $\langle \hat{\mathbf{s}}(\mathbf{r}) \rangle^\mathrm{ad}_t$. This result was also re-derived microscopically in Ref.~\cite{Cheng2013} to reveal some of the limiting assumptions behind the phenomenological expression, such as that local equilibrium can be defined and $\mathbf{M}_i(t)$ has to be nearly constant in time during time scale set by $\hbar/J_\mathrm{sd}$. 

{\em Models and methods}.---The LMMs hosted by FM nanowire in Fig.~\ref{fig:fig1} are described by a classical Hamiltonian
\begin{equation}\label{eq:classicalh}
{\mathcal H}  =  -J \sum_{\langle i j \rangle} 	\mathbf{M}_{i}\cdot \mathbf{M}_{j}	- K\sum_{i} {\left(M_{i}^{z}\right)}^2	+ D\sum_{i}	{\left(M_{i}^{y}\right)}^2, 
\end{equation}
where \mbox{$J = 0.1\ {\rm eV}$} is the Heisenberg exchange coupling between the nearest-neighbor (NN) LMMs; \mbox{$K = a^2 J/w^2$} is  magnetic anisotropy along the $z$-axis; and \mbox{$D=0.03$ meV} is demagnetization field. The spatial profile of a N\'{e}el DW in equilibrium, or in the presence of injected current but kept static as in Figs.~\ref{fig:fig2}, ~\ref{fig:fig4}(c) and ~\ref{fig:fig4}(d), is given by $\mathbf{M}_i=\big(\sech[(X_{\rm DW}-i)/w], 0, \tanh[(X_{\rm DW}-i)/w] \big)$, where $X_{\rm DW}$ is the center of the DW.

The same sites also host conduction electrons described by a quantum tight-binding Hamiltonian 
\begin{equation}\label{eq:quantumh}
\hat{H}(t) = -\gamma \sum_{\langle ij \rangle}  \hat{c}_{i}^\dagger\hat{c}_j - J_\mathrm{sd} \sum_{i}\hat{c}_i^\dagger\boldsymbol{\sigma} \cdot \bold{M}_i(t) \hat{c}_i, 
\end{equation}
where \mbox{$\hat{c}_i^\dagger = (\hat{c}_{i\uparrow}^\dagger,\hat{c}_{i\downarrow}^\dagger)$}  is a row vector containing operators $\hat{c}_{i\sigma}^\dagger$ which create an electron of spin $\sigma=\uparrow,\downarrow$ at the site $i$; $\hat{c}_i$ is a column vector that contains the corresponding  annihilation operators; \mbox{$\gamma=1$ eV} is the NN hopping; and \mbox{$J_{\rm sd}=0.1$ eV}~\cite{Cooper1967}. The FM nanowire is attached to semi-infinite NM leads, modeled by the first term alone in  $\hat{H}(t)$ [Eq.~\eqref{eq:quantumh}], which terminate into macroscopic reservoirs at infinity. The Fermi energy of the reservoirs is set at \mbox{$E_F=0$~eV} in Fig.~\ref{fig:fig2} or \mbox{$E_F=-1.95$~eV} in Figs.~\ref{fig:fig3} and ~\ref{fig:fig4}. Time dependence of $\mathbf{M}_i(t)$ entering into $\hat{H}(t)$ is obtained by solving a system of LLG equations~\cite{Evans2014}
\begin{eqnarray}\label{eq:llglattice}
	\partial_t \mathbf{M}_i & = & -g_0 \mathbf{M}_i \times \mathbf{B}^\mathrm{eff}_i + \lambda \mathbf{M}_i \times \partial_t \mathbf{M}_i \nonumber \\
	&& + \frac{g_0}{\mu_M}\left( \mathbf{T}_i^\mathrm{ad} + \mathbf{T}_i^\mathrm{nad} \right),  
\end{eqnarray}
where \mbox{$\mathbf{B}^{\rm eff}_i = - \frac{1}{\mu_M} \partial \mathcal{H} /\partial \mathbf{M}_{i}$} is obtained from $\mathcal{H}$ in Eq.~\eqref{eq:classicalh}; $\mu_M$ is the magnitude of LMMs~\cite{Evans2014}; and Gilbert damping is chosen as $\lambda=0.01$ as typically measured~\cite{Weindler2014} in metallic ferromagnets. In the case of current-driven motion of DW analyzed in Figs.~\ref{fig:fig3}, ~\ref{fig:fig4}(a) and ~\ref{fig:fig4}(b), we self-consistently couple Eq.~\eqref{eq:llglattice} to calculations of  $\mathbf{T}_i^\mathrm{nad}$ and $\mathbf{T}_i^\mathrm{ad}$ via Eqs.~\eqref{eq:nadstt} and ~\eqref{eq:adstt}, respectively, in terms of nonequilibrium density matrices obtained from TDNEGF formalism, as explained in Refs.~\cite{Petrovic2018,Bajpai2019a}. Such quantum-classical TDNEGF+LLG approach uses time step \mbox{$\delta t=0.1$ fs} in both quantum [i.e., TDNEGF computation of ${\bm \rho}^{\rm neq}(t) = \hbar \mathbf{G}^<(t,t)/i$] and classical [i.e., LLG Eq.~\eqref{eq:llg}] parts of the self-consistent loop~\cite{Petrovic2018,Bajpai2019a}.

{\em Results and discussion}.---We warm up by first establishing specific range of  $J_\mathrm{sd}w/\gamma a$---which is the 
re-written form of Eq.~\eqref{eq:clean} with $v_F \propto \gamma$ and $a$ as the lattice spacing, as pertinent for tight-binding lattice of (clean) FM nanowire in Fig.~\ref{fig:fig1}---that ensures putative adiabatic limit~\cite{Tatara2019,Tatara2000,Gopar2004} with {\em no reflection} of electronic spin from the DW. Note that we vary $J_\mathrm{sd} w/\gamma a$ in Figs.~\ref{fig:fig2} and ~\ref{fig:fig3} by either changing $J_\mathrm{sd}$ at fixed $w$ (circles) or by changing $w$ at fixed $J_\mathrm{sd}$ (triangles), so that overlap of these two sets of data confirms that $J_\mathrm{sd}w/\gamma a$ is indeed relevant parameter for adiabaticity. For this purpose, we analyze in Fig.~\ref{fig:fig2} how unpolarized charge current becomes spin polarized, and how its polarization then adjusts to DW profile by computing local (or bond) spin currents~\cite{Nikolic2006} (instead of previously used Landauer spin-resolved transmission functions~\cite{Tatara2000,Gopar2004}) $I^{S_\alpha}_{i \rightarrow i+1}(t)$ from site $i$ to site $i+1$ before ($i=2$) and after  ($i=38$) {\em static} (i.e., not allowed to move) DW in Fig.~\ref{fig:fig1}. The unpolarized charge current is injected by time-dependent bias voltage \mbox{$V_b(t) = V_\mathrm{max} \big[\tanh((t-t_0)/\tau)+1 \big]$}, where \mbox{$V_\mathrm{max}=0.05$ eV}, \mbox{$t_0 = 25$ fs}, and \mbox{$\tau = 10$ fs}. 

Negative $I^{S_z}_{2 \rightarrow 3}(t)$ for small $J_\mathrm{sd}$ [Figs.~\ref{fig:fig2}(c) and ~\ref{fig:fig2}(d)] means that most of spins are reflected [our convention is that positive $I^{S_\alpha}_{i \rightarrow i+1}(t)$  means spin-up along the $\alpha$-axis moves from site $i$ to $i+1$], with $I^{S_z}_{2 \rightarrow 3}(t) \rightarrow 0$ around \mbox{$J_\mathrm{sd} w/\gamma a \simeq 1$}. Such reflection is nonadiabatic effect, where previous studies have explored thereby generated force~\cite{Tatara2019} on DW and its electrical resistance~\cite{Levy1997,Tatara1997,Tatara2000,Simanek2001,Simanek2005,Berger2007,Gopar2004,Tatara2019}. Our time-dependent QT approach also reveals [Fig.~\ref{fig:fig2}(a)] that $I^{S_z}_{38 \rightarrow 39}(t)$ is not immediately negative, as expected for spin-down (along the $z$-axis) moving in the direction of the positive $x$-axis in Fig.~\ref{fig:fig1}, but it takes about \mbox{$\simeq 50$ fs} for that to happen [dashed red curve in Fig.~\ref{fig:fig2}(a)]. Figure~\ref{fig:fig2}(b) shows that adiabatic limit, in which $I^{S_z}_{38 \rightarrow 39}(t)/I^{S_z}_{2 \rightarrow 3}(t) \rightarrow -1$ due to electronic 
spin along the $z$-axis smoothly changing orientation from up to down to surrender to surrender~\cite{Waintal2004,Tatara2019} a quantum $\hbar$ of angular momentum on the DW, is established for $J_\mathrm{sd}w/\gamma a \gtrsim 1.5$. In this limit, $\langle \hat{\mathbf{s}}_i \rangle^\mathrm{neq}(t)$ tracks [Fig.~\ref{fig:fig2}(d)] adiabatic spin density $\langle \hat{\mathbf{s}}_i \rangle^\mathrm{ad}_t$, but $\langle \hat{\mathbf{s}}_i \rangle^\mathrm{ad}_t$ must continue to mistrack $\mathbf{M}_i$ [Fig.~\ref{fig:fig2}(c)] so that $\int \! dt\, J_\mathrm{sd} \langle \hat{\mathbf{s}}_i \rangle^\mathrm{ad}_t \times \mathbf{M}_i = \hbar$ (assuming single-electron current pulse injection~\cite{Keeling2006,Suresh2020}).

The magnitude of nonadiabatic and adiabatic STT across FM nanowire hosting current-driven {\em moving} DW is show in Fig.~\ref{fig:fig3}(d), with their maximum 
values~\cite{Xiao2006} analyzed in Fig.~\ref{fig:fig3}(a)--(c). Figure~\ref{fig:fig3}(a) demonstrates that nonadiabatic STT does not decay to zero in adiabatic limit (as deciphered from Fig.~\ref{fig:fig2}) of wide DW walls, but instead saturates at an asymptotic value. This result is in full accord with experiments~\cite{Boulle2008,Burrowes2010,Heyne2010} where relative insensitivity of nonadiabatic STT to DW width was observed. In addition, phenomenological nonadiabatic STT is out-of-DW-plane~\cite{Zhang2004} (see also Fig. 19 in Ref.~\cite{Tatara2019}), which has already been questioned by Ref.~\cite{Xiao2006} which finds additional in-plane component  that is reproduced by our Fig.~\ref{fig:fig4}(a). However, our out-of-plane (i.e., along the $y$-axis) component in Fig.~\ref{fig:fig4}(b) remains finite with increasing DW width, while that of Ref.~\cite{Xiao2006} decays exponentially with $w$. 

We note, however, that STT in Ref.~\cite{Xiao2006} was computed for static DW using time-independent QT. So, in Fig.~\ref{fig:fig4}(c) we also employ static DW to demonstrate that out-of-plane nonadiabatic STT does decay in Figs.~\ref{fig:fig4}(d), in contrast to STT on dynamic DW in Fig.~\ref{fig:fig4}(a). Interestingly, in the case of static DW nonadiabatic STT in Fig.~\ref{fig:fig4}(c) is {\em nonzero} outside of DW region, which was pointed out in Ref.~\cite{Xiao2006} to be the signature of {\em nonlocality} of nonadiabatic STT. But such effect is an artifact of using static DW in calculations, and it disappears in full TDNEGF+LLG self-consistent calculations of Fig.~\ref{fig:fig4}(a).

{\em Conclusions and outlook}.---Using numerically {\em exact} time-dependent QT~\cite{Gaury2014,Petrovic2018,Bajpai2019a}, we demonstrate that wide DWs harbor {\em adiabatic} spin transport, in which no spin reflection from {\em static} DW takes place [Fig.~\ref{fig:fig2}(b)] as analyzed previosuly~\cite{Tatara2000,Tatara2019,Gopar2004}, as well as {\em nonadiabaticity} [Figs.~\ref{fig:fig3} and ~\ref{fig:fig4}] of microscopically extracted nonequilibrium spin density [Eq.~\eqref{eq:snonadiabatic}] and thereby generated STT [Eq.~\eqref{eq:nadstt}] on  {\em moving} DW. These findings are in full accord with experiments~\cite{Boulle2008,Burrowes2010,Heyne2010}. The origin of discrepancy between them and prior analysis~\cite{Zhang2004,Tatara2019,Cheng2013}, providing widely used~\cite{Boulle2008,Burrowes2010,Heyne2010,Alejos2021} in micromagnetics phenomenological expression [Eq.~\eqref{eq:llg}] for nonadiabatic STT that is  exponentially suppressed in wide DWs~\cite{Xiao2006}, can  be traced [Fig.~\ref{fig:fig4}] to the usage of {\em static}~\cite{Xiao2006} or {\em effectively static}~\cite{Cheng2013} [i.e., $\mathbf{M}_i(t)$ is assumed to be nearly constant in time on the time scale $\hbar/J_\mathrm{sd}$]  DWs,  as well as general expectation that electrons can be completely ``integrated out''~\cite{Zhang2004,Onoda2006,Bhattacharjee2012} to arrive at an effective formula. Such formula, plugged into the LLG Eq.~\eqref{eq:llg}, then makes  possible describing current-driven DW motion using solely classical micromagnetics~\cite{Thiaville2005,Alejos2021}. 

However, it has been demonstrated recently~\cite{Bajpai2019a,Sayad2015} that such strategy can work only when $J_\mathrm{sd}$ is weak and can be treated perturbatively~\cite{Onoda2006,Bhattacharjee2012}. In other regimes, self-consistent~\cite{Petrovic2018,Bajpai2019a,Sayad2015,Hurst2020}  coupling 
of quantum dynamics of electrons to LLG equation for LMMs can lead to a plethora of new effects in the classical dynamics of LMMs comprising DW---such as nonlocal-in-time response kernel which generates both additional damping~\cite{Bajpai2019a,Bhattacharjee2012,Petrovic2021} and inertia (even in clean system)~\cite{Bajpai2019a,Bhattacharjee2012,Hurst2020}---whose intricacy can hardly be captured by a simple expression as in Eq.~\eqref{eq:llg}.

\begin{acknowledgments}
	This work was supported by the US National Science Foundation (NSF) Grant No. ECCS 1922689. 
\end{acknowledgments}


\begin{thebibliography}{10}
	
	\bibitem{Tatara2008}
	G.~Tatara, H.~Kohno, and J.~Shibata, Microscopic approach to current-driven domain wall dynamics, Phys. Rep. {\bf 468}, 213 (2008).
	
	\bibitem{Tatara2019}
	G. Tatara, Effective gauge field theory of spintronics, Physica E {\bf 106}, 208 (2019).
	
	\bibitem{Boulle2008}
	O. Boulle, J. Kimling, P. Warnicke, M. Kl\"{a}ui, U. R\"{u}diger, G. Malinowski, H. J. M. Swagten, B. Koopmans, C. Ulysse, and G. Faini, Nonadiabatic spin transfer torque in high anisotropy magnetic nanowires with narrow domain walls, Phys. Rev. Lett. {\bf 101}, 216601 (2008).
	
	\bibitem{Burrowes2010}
	C. Burrowes {\em et al.}, Non-adiabatic spin-torques in narrow magnetic domain walls, Nat. Phys. {\bf 6}, 17 (2010).
	
	\bibitem{Akosa2017}
	C. A. Akosa, P. B. Ndiaye, and A. Manchon, Intrinsic nonadiabatic topological torque in magnetic skyrmions and vortices, Phys. Rev. B {\bf 95}, 054434  (2017).
	
	\bibitem{Heyne2010}
	L. Heyne, J. Rhensius, D. Ilgaz, A. Bisig, U. R\"{u}diger, M. Kl\"{a}ui, L. Joly, F. Nolting, L. J. Heyderman, J. U. Thiele, and F. Kronast,  Direct determination of large spin-torque nonadiabaticity in vortex core dynamics, Phys. Rev. Lett. {\bf 105}, 187203 (2010).
	
	\bibitem{Pollard2012}
	S. D. Pollard, L. Huang, K. S. Buchanan, D. A. Arena, and Y. Zhu, Direct dynamic imaging of non-adiabatic spin torque effects, Nat. Commun. {\bf 3}, 1028 (2012).
	
	\bibitem{Levy1997}
	P. M. Levy and S. Zhang, Resistivity due to domain wall scattering, Phys. Rev. Lett. {\bf 79}, 5110 (1997).
	
	\bibitem{Zhang2004}
	S. Zhang and Z. Li, Roles of nonequilibrium conduction electrons on the magnetization dynamics of ferromagnets, Phys. Rev. Lett. {\bf 93}, 127204 (2004).
	
	\bibitem{Barnes2005}
	S. E. Barnes and S. Maekawa, Current-spin coupling for ferromagnetic domain walls in fine wires, Phys. Rev. Lett. {\bf 95}, 107204 (2005).
	
	\bibitem{Gregg1996}
	J. F. Gregg, W. Allen, K. Ounadjela, M. Viret, M. Hehn, S. M. Thompson, and J. M. D. Coey, Giant magnetoresistive effects in a single element magnetic thin 
	film, Phys. Rev. Lett. {\bf 77}, 1580 (1996).
	
	\bibitem{Berger1978}
	L. Berger, Low‐field magnetoresistance and domain drag in ferromagnets, J. Appl. Phys. {\bf 49}, 2156 (1978).
	
	\bibitem{Aharonov1992}
	Y. Aharonov and A. Stern, Origin of the geometric forces accompanying Berry's geometric potentials, Phys. Rev. Lett. {\bf 69}, 3593 (1992).
	
	\bibitem{Xiao2006}
	J. Xiao, M. D. Stiles, and A. Zangwill, Spin-transfer torque for continuously variable magnetization, Phys. Rev. B {\bf 73}, 054428 (2006).
	
	\bibitem{Kim2017a}
	K.-J. Kim, Y.~Yoshimura, and T.~Ono, Current-driven magnetic domain wall motion and its real-time detection, Jap. J. Appl. Phys., {\bf 56} 0802A4 (2017).
	
	\bibitem{Parkin2015}
	S.~Parkin and S.-H. Yang, Memory on the racetrack, Nat. Nanotech. {\bf 10} 195 (2015).
	
	\bibitem{Allwood2002}
	D.~A. Allwood, G.~Xiong, M.~D. Cooke, C.~C. Faulkner, D.~Atkinson, N.~Vernier, and R.~P. Cowburn, Submicrometer ferromagnetic not gate and shift register,  Science {\bf 296}, 2003 (2002).
	
	\bibitem{Grollier2016}
	J.~Grollier, D.~Querlioz, and M.~D. Stiles, Spintronic nanodevices for bioinspired computing, Proc. IEEE {\bf 104}, 2024 (2016).
	
	\bibitem{Cooper1967}
	R. L. Cooper and E. A. Uehling, Ferromagnetic resonance and spin diffusion in supermalloy, Phys. Rev. {\bf 164}, 662 (1967).
	
	\bibitem{Waintal2004}
	X. Waintal and M. Viret, Current-induced distortion of a magnetic domain wall, EPL (Europhysics Letters) {\bf 65}, 427 (2004).
	
	\bibitem{Ban2009}
	Y. Ban and G. Tatara, Spin-transfer torque in disordered weak ferromagnets, Phys. Rev. B {\bf 80}, 184406 (2009).	
	
	\bibitem{Berger2007}
	L. Berger, Relation between damping, current-induced torques, and wall resistance for domain walls in magnetic nanowires, Phys. Rev. B {\bf 75}, 174401 (2007).
	
	\bibitem{Tatara1997}
	G. Tatara and H. Fukuyama, Resistivity due to a domain wall in ferromagnetic metal, Phys. Rev. Lett. {\bf 78}, 3773 (1997).
	
	\bibitem{Tatara2000}
	G. Tatara, Domain wall resistance based on Landauer's formula, J. Phys. Soc. Jpn. {\bf 69}, 2969 (2000).
	
	\bibitem{Simanek2001}
	E. \v{S}im\'{a}nek, Spin accumulation and resistance due to a domain wall, Phys. Rev. B 63, 224412 (2001).
	
	\bibitem{Simanek2005}
	E. \v{S}im\'{a}nek and A. Rebei, Spin transport and resistance due to a Bloch wall, Phys. Rev. B {\bf 71}, 172405 (2005).
	
	\bibitem{Gopar2004}
	V. A. Gopar, D. Weinmann, R. A. Jalabert, and R. L. Stamps, Electronic transport through domain walls in ferromagnetic nanowires: Coexistence of adiabatic and nonadiabatic spin dynamics, Phys. Rev. B {\bf 69}, 014426 (2004).
	
	\bibitem{Thiaville2005}
	A. Thiaville, Y. Nakatani, J. Miltat, and Y. Suzuki, Micromagnetic understanding of current-driven domain wall motion in patterned nanowires, EPL (Europhysics Letters) {\bf 69}, 990 (2005).
	
	\bibitem{Ralph2008}
	D.~Ralph and M.~Stiles, Spin transfer torques, J. Magn. Mater. {\bf 320}, 1190 (2008).
	
	\bibitem{Bazaliy1998}
	Ya. B. Bazaliy, B. A. Jones, and S.-C. Zhang, Modification of the Landau-Lifshitz equation in the presence of a spin-polarized current in colossal- and giant-magnetoresistive materials, Phys. Rev. B {\bf 57}, R3213 (1998).
	
	\bibitem{Tatara2004}
	G. Tatara and H. Kohno, Theory of current-driven domain wall motion: Spin transfer versus momentum transfer, Phys. Rev. Lett. {\bf 92}, 086601 (2004).	
	
	\bibitem{Fernandez-Rossier2004}
	J. Fernańdez-Rossier, M. Braun, A. S. N\'{u}\~{n}ez, and A. H. MacDonald, Influence of a uniform current on collective magnetization dynamics in a ferromagnetic metal, Phys. Rev. B {\bf 69}, 174412 (2004).
	
	\bibitem{Cheng2013}
	R. Cheng and Q. Niu, Microscopic derivation of spin-transfer torque in ferromagnets, Phys. Rev. B {\bf 88}, 024422 (2013).
	
	\bibitem{Evans2014}
	R.~F.~L. Evans, W.~J. Fan, P.~Chureemart, T.~A. Ostler, M.~O.~A. Ellis, and R.~W. Chantrell, Atomistic spin model simulations of magnetic
	nanomaterials, J. Phys.: Condens. Matter {\bf 26}, 103202 (2014).
	
	\bibitem{Duine2007}
	R. A. Duine, A. S.  N\'{u}\~{n}ez, J. Sinova, and A. H. MacDonald, Functional Keldysh theory of spin torques, Phys. Rev. B {\bf 75}, 214420 (2007).
	
	\bibitem{Kishine2010}
	J.-I. Kishine and A. S. Ovchinnikov, Adiabatic and nonadiabatic spin-transfer torques in the current-driven magnetic domain wall motion, Phys. Rev. B {\bf 81}, 134405 (2010).
	
	\bibitem{Mondal2018}
    R. Mondal, M. Berritta, and P. M. Oppeneer, Unified theory of magnetization dynamics with relativistic and nonrelativistic spin torques, Phys. Rev. B {\bf 98}, 214429 (2018).
	
	\bibitem{Kohno2006}
	H. Kohno, G. Tatara, and J. Shibata, Microscopic calculation of spin torques in disordered ferromagnets, J. Phys. Soc. Japan {\bf 75}, 113706 (2006).
	
	\bibitem{Tatara2008a}
	G. Tatara and P. Entel, Calculation of current-induced torque from spin continuity equation, Phys. Rev. B {\bf 78}, 064429 (2008).
	
	\bibitem{Nguyen2007}
	A. K. Nguyen, H. J. Skadsem, and A. Brataas, Giant current-driven domain wall mobility in (Ga,Mn)As, Phys. Rev. Lett. 98, 146602 (2007).
	
	\bibitem{Garate2009}
	I. Garate, K. Gilmore, M. D. Stiles, and A. H. MacDonald, Nonadiabatic spin-transfer torque in real materials, Phys. Rev. B {\bf 79}, 104416 (2009).
	
	\bibitem{Nikolic2006}
	B. K. Nikoli\'{c}, L. P. Z\^{a}rbo, and S. Souma, Imaging mesoscopic spin Hall fow: Spatial distribution of local spin currents and spin densities in and out of multiterminal spin-orbit coupled semiconductor nanostructures, Phys. Rev. B {\bf 73}, 075303 (2006).
	
	\bibitem{Petrovic2018}
	M.~D. Petrovi\'{c}, B.~S. Popescu, U.~Bajpai, P.~Plech\'{a}\v{c}, and B.~K. Nikoli\'{c}, Spin and charge pumping by a steady or pulse-current-driven magnetic domain wall: A self-consistent multiscale time-dependent quantum-classical hybrid approach, Phys. Rev. Applied {\bf 10}, 054038 (2018).
	
	\bibitem{Bajpai2019a}
	U.~Bajpai and B.~K. Nikoli\'{c}, Time-retarded damping and magnetic inertia in the Landau-Lifshitz-Gilbert equation self-consistently coupled to electronic time-dependent nonequilibrium Green functions, Phys. Rev. B {\bf 99}, 134409 (2019).	
	
	\bibitem{Bajpai2020}
	U. Bajpai and B. K. Nikoli\'{c}, Spintronics meets nonadiabatic molecular dynamics: Geometric spin torque and damping on dynamical classical magnetic texture due to an electronic open quantum system, Phys. Rev. Lett. {\bf 125}, 187202 (2020).
	
	\bibitem{Petrovic2021}
	M.~D. Petrovi\'{c},  U.~Bajpai, P.~Plech\'{a}\v{c}, and B.~K. Nikoli\'{c}, Annihilation of topological solitons in magnetism with spin wave burst finale: The role of nonequilibrium electrons causing nonlocal damping and spin pumping over ultrabroadband frequency range, Phys. Rev. B {\bf 104}, L020407 (2021). 
	
	\bibitem{Gaury2014}
	B.~Gaury, J.~Weston, M.~Santin, M.~Houzet, C.~Groth, and X.~Waintal, Numerical simulations of time-resolved quantum electronics, Phys. Rep. {\bf 534}, 1 (2014).
	
	\bibitem{Popescu2016}
	B.~S. Popescu and A.~Croy, Efficient auxiliary-mode approach for time-dependent nanoelectronics, New J. Phys. {\bf 18}, 093044 (2016).	
	
	\bibitem{Stahl2017}
	C. Stahl and M. Potthoff, Anomalous spin precession under a geometrical torque, Phys. Rev. Lett. {\bf 119},  227203  (2017).
	
	\bibitem{Capelle2001}
	K. Capelle, G. Vignale, and B. L. Gy\"{o}rffy, Spin currents and spin dynamics in time-dependent density-functional theory, Phys. Rev. Lett. {\bf 87}, 206403 (2001). 
	
	\bibitem{Bode2011}
	N. Bode, S.~V. Kusminskiy, R. Egger, and F. von Oppen, Scattering theory of current-induced forces in mesoscopic systems, Phys. Rev. Lett. {\bf 107},  036804  (2011).
	
	\bibitem{Thomas2012a}
	M. Thomas, T. Karzig, S.~V. Kusminskiy, G. Zar\'and, and F. von Oppen, Scattering theory of adiabatic reaction forces due to out-of-equilibrium quantum environments, Phys. Rev. B {\bf 86},  195419  (2012).
	
	\bibitem{Mahfouzi2016}
	F. Mahfouzi, B. K. Nikoli\'{c}, and N. Kioussis, Antidamping spin-orbit torque driven by spin-flip reflection mechanism on the surface of a topological insulator: A time-dependent nonequilibrium Green function approach, Phys. Rev. B {\bf 93}, 115419 (2016). 
	
	\bibitem{Stefanucci2013}
	G.~Stefanucci and R.~van Leeuwen, \emph{Nonequilibrium Many-Body Theory of Quantum Systems: A Modern Introduction} (Cambridge University Press, Cambridge, 2013).
	
	\bibitem{Weindler2014}
	T. Weindler, H. G. Bauer, R. Islinger, B. Boehm, J.-Y. Chauleau, and C. H. Back, Magnetic damping: Domain wall dynamics versus local ferromagnetic resonance, 
	Phys. Rev. Lett. {\bf 113}, 237204 (2014).
	
	\bibitem{Keeling2006}
	J. Keeling, I. Klich, and L. Levitov, Minimal excitation states of electrons in one-dimensional wires, Phys. Rev. Lett. {\bf 97}, 116403 (2006).
	
	\bibitem{Suresh2020}
	A. Suresh, U. Bajpai, and B. K. Nikoli\'{c}, Magnon-driven chiral charge and spin pumping and electron-magnon scattering from time-dependent quantum transport combined with classical atomistic spin dynamics, Phys. Rev. B {\bf 101}, 214412 (2020).
		
	\bibitem{Alejos2021}
	O. Alejos, V. Raposo, and E. Mart\'{i}nez, Domain wall motion in magnetic nanostrips, in {\em Materials Science and Technology}, eds. R. W. Cahn, P. Haasen and E. J. Kramer (Wiley, Hoboken, 2021).
	
	\bibitem{Onoda2006}
	M. Onoda and N. Nagaosa, Dynamics of localized spins coupled to the conduction electrons with charge and spin currents, Phys. Rev. Lett. {\bf 96}, 066603 (2006).
	
	\bibitem{Bhattacharjee2012}
	S. Bhattacharjee, L. Nordstr\"{o}m, and J. Fransson, Atomistic spin dynamic method with both damping and moment of inertia effects included from first principles Phys. Rev. Lett. {\bf 108}, 057204 (2012).
	
	\bibitem{Sayad2015}
	M.~Sayad and M.~Potthoff, Spin dynamics and relaxation in the classical-spin Kondo-impurity model beyond the Landau-Lifschitz-Gilbert equation, New J. Phys. {\bf 17}, 113058 (2015). 	
	
	\bibitem{Hurst2020}
	H. M. Hurst, V. Galitski, and T. T. Heikkil\"{a}, Electron-induced massive dynamics of magnetic domain walls, Phys. Rev. B {\bf 101}, 054407 (2020).

		
\end{thebibliography}

\end{document}